\documentclass[a4paper,10pt,twocolumn]{revtex4}           
\usepackage{graphicx}                           
\usepackage[ansinew]{inputenc}                  
\usepackage{amsfonts}                           
\usepackage[normalem]{ulem}                     
\usepackage{amsmath}                            

\newcommand{\ident}{ \ {\rm l\hspace{-5.55pt}\mathbf{1}}}

\begin{document}

\author{J. G. Oliveira Junior$^{1,2}$\footnote{Electronic
address: zgeraldo@ufrb.edu.br}, J. G. Peixoto de
Faria$^{3}$\footnote{ Electronic address: jgpfaria@des.cefetmg.br},
M. C. Nemes$^{2}$\footnote{Electronic address:
carolina@fisica.ufmg.br}}

\affiliation{$^1$Centro de Forma\c c\~ao de Professores,
Universidade Federal do
Rec\^oncavo da Bahia, 45.300-000, Amargosa, BA, Brazil\\
$^2$Departamento de F\'{\i}sica - CP 702 - Universidade
Federal de Minas Gerais - 30123-970 - Belo Horizonte - MG - Brazil\\
$^3$Departamento de Física e Matemática - Centro Federal de Educa\c
c\~ao Tecnol\'ogica de Minas Gerais - 30510-000 - Belo Horizonte -
MG - Brazil}

\title{Degree of information accessibility: a precursor of the quantum--classical transition}


%

\date{\today }

\maketitle

One of the fundamental problems with the interpretation of Quantum
Mechanics, according to Bohr \cite{bohr}, is the fact that
``\emph{our usual description of physical phenomena is based
entirely on the idea that the phenomena concerned may be observed
without disturbing them appreciably}". Furthermore in his articles
\cite{bohr,bohr2} discussing the subject Bohr argues that the action
of the probe will be affected by the system and inasmuch the system
will be affected by the probe. Specifically in {\it Gedanken}
experiments he tests the wave--particle duality of the system and
implicitly assumes that the probe is also a quantum system. A
universal character can only be attributed to Quantum Mechanics
provided a complementarity relation is also valid for the probe. As
a consequence the state system--probe becomes entangled as
extensively discussed by N. Bohr, W. Heisenberg, A. Einstein, E.
Schr\"odinger, de Broglie and W. Pauli among others starting in the
famous Solvay conference \cite{einstein}. Soon after entanglement
has been brought to discussion \cite{scho,epr} and since then it has
always been present in several contexts \cite{emaranhamento}. In the
past fourty years complementarity tests have been proposed as, {\it
e.g.}, using double slit experiments \cite{wootters}, quantum eraser
\cite{scully1,nature,steve,zai}, cavity interferometry
\cite{scully2, haroche}, and experiments involving which--way
detection \cite{englert,rempe,rempe2,mahalo}, to quote some.
However, much less attention has been paid to the study of an
arbitrary probe system. In the present contribution, we fill in this
gap and show that the key ingredient for the quantum--classical
transition is not necessarily the information generated by the
system--probe interaction but rather by its accessibility. Our
results have been successfully tested in the interferometric
experiment by \cite{haroche}. Our results also allow for a simple
physical interpretation of the physics of Ramsey Zones
\cite{ramsey1} where one photon (average) interacts with a two level
atom in a classical manner, {\it i.e.}, no entanglement is
generated.

\begin{figure}[!htb]
\centering
  \includegraphics[scale=0.300,angle=00]{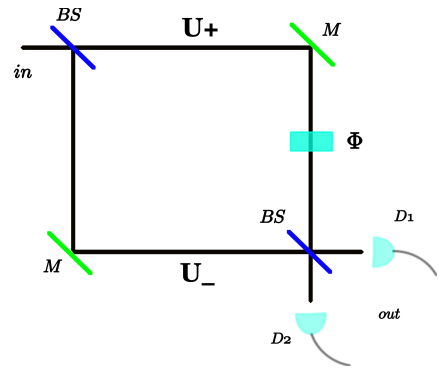}\\
  \caption{{\small
Mach--Zehnder interferometer with perfect beam splitters ($BS$) and
mirrors ($M$). The vector $|1\rangle$ ($|0\rangle$) represent the
situation where the particle goes through the upper (lower) arm,
respectively. After the first beam splitter the particle interacts
with the probe. It is then reflected by a mirror and acquires a
relative phase ($\Phi$).  In the sequel it is detected in $D_{1}$ or
$D_{2}$ and will exhibit an interference pattern with sinusoidal
modulation as a function of $\Phi$. The action of the $BS$, of the
probe and of the acumulated relative phase between in the arms will
be respectively modelled by the following unitary operators
$U_{B}=(\ident-i\sigma_y)\otimes\ident/\sqrt{2} $,
$U_{P}=|1\rangle\langle 1|\otimes U_{+}+|0\rangle\langle 0|\otimes
U_{-}$ e $U_{F}=(e^{i\Phi}|1\rangle\langle 1|+|0\rangle\langle
0|)\otimes\ident $ where $U_{\pm}|m\rangle=|m_{\pm}\rangle$,
$U_{\pm}^{}U_{\pm}^{\dag}=\ident$ and
$\sigma_y=\bigl(-i|1\rangle\langle0|+i|0\rangle\langle1|\bigr)$.
}}\label{mach}
\end{figure}

Let us consider a Mach--Zehnder interferometer (see \cite{bush} for
a revision) with a probe system capable of obtaining and storing the
which--way information provided by a particle sent through it, as
shown in Figure \ref{mach}.%
In the process of obtaining and storing which--way information, the
probe and particle will entangle. This results in a kind of
Einstein--Podolsky--Rosen \cite{epr} pair. The configuration is that
of a bipartite system defined in the Hilbert space
$\mathcal{H}^{(1)}\otimes\mathcal{H}^{(2)}$, where the subsystem in
$\mathcal{H}^{(1)}$ has two levels and is entangled to another
system in $\mathcal{H}^{(2)}$. In general one can describe in the
Schmidt base \cite{peres}
\begin{equation}\label{ximiti}
|\Psi\rangle=\sum_{i=1}^{2}\sqrt{\lambda_i\,}|v^{(1)}_{i}\rangle|v^{(2)}_{i}\rangle
\end{equation}
where $\{|v^{(1)}_{i}\rangle\}$ ($\{|v^{(2)}_{i}\rangle\}$) are
orthonormal bases in $\mathcal{H}^{(1)}$ ($\mathcal{H}^{(2)}$),
respectively, and the $\sqrt{\lambda_i\,}$ are the Schmidt
coefficients. The state of the system in $\mathcal{H}^{(1)}$ will be
$\rho_{(1)}=\sum_{i=1}^{2}\lambda_i|v^{(1)}_{i}\rangle\langle
v^{(1)}_{i}|$\,. Since $\lambda_1+\lambda_2=1$ it follows
immediately that $(\lambda_{1}-\lambda_{2})^{2}+4\det\rho^{(1)}=1$,
where $4\det\rho^{(1)}$ is the concurrence $C^{}_{(1,2)}$ squared
\cite{wootters2} between particle--probe. The system in
$\mathcal{H}^{(1)}$ is a two level system and therefore can be
written in terms of Pauli matrices
$\{\sigma_{-},\sigma^{\dag}_{-},\sigma_{z}\}$. The equation above
becomes $
\langle\sigma^{(1)}_{z}\rangle^{2}+4\,|\langle\sigma^{(1)}_{-}\rangle|^{2}+C^{2}_{(1,2)}=1
$, where
$\langle\sigma^{(1)}_{k}\rangle=\mathrm{tr}(\rho_{(1)}\sigma^{}_{k})$.
Note that as we identify the $\sigma^{(1)}_{z}$ eigenstates with the
arms of the interferometer, $|\langle \sigma^{(1)}_{z}\rangle|$
represents the probability of finding the particle in one of the
arms, usually called {\it predictability} $\mathcal{P}_{(1)}$
\cite{englert,bergou2}. On the other hand, the term
$|\langle\sigma^{(1)}_{-}\rangle|$ represents the magnitude of the
coherence between the interferometers arms. The interference fringes
will exhibit a {\it visibility}
$\mathcal{V}_1=2|\langle\sigma^{(1)}_{-}\rangle|$ \cite{bergou2}.
Having these two results it follows that
$\mathcal{S}_{(1)}^{2}+C^{2}_{(1,2)}=1$, result first found in
\cite{bergou2}, where
$\mathcal{S}_{(1)}^{2}=\mathcal{P}_{(1)}^{2}+\mathcal{V}_{(1)}^{2}$
represents wave--particle duality. Given the symmetry of the Schmidt
decomposition the wave--particle duality of the other system will be
the same,
\begin{equation}\label{igualdade_dualidade}
\mathcal{S}_{(2)}^{2}=\mathcal{S}_{(1)}^{2} \ \ \ \ .
\end{equation}

In our problem the second system is the probe. Let us consider that
the particle has been prepared in the state
$(|1\rangle\pm|0\rangle)/\sqrt{2}$ and sent through the
interferometer. The probe, prepared in $|m\rangle$, evolves to
$|m_\mp\rangle$. We may at this point ask for the probability that a
measurement of the state of the probe will yield $|m_\pm\rangle$.
This probability will be $\mathcal{O}=|\langle m_-|m_+\rangle|^2$,
for both cases and may be interpreted as the imperfection of the
probe. As a consequence we may define $\mathcal{Q}=1-|\langle
m_-|m_+\rangle|^2$ a quantity which represents the quality of the
probe. The quantity $\,\mathcal{D}=\sqrt{\mathcal{Q}}$\, is a
quantitative measure of the distinguishability of the probe states
\cite{englert}, {\it i.e.}, of the which--way information available
in the probe system. In this notation we can say that a perfect
probe with $\mathcal{D}=1$ is the one which makes the which--way
information completely available. Thus a perfect probe
($\mathcal{Q}=1$) will have completely distinguishable probe states
($|\langle m_-|m_+\rangle|=0$) and the particle state will be a
statistical mixture showing thus no interference pattern. On the
other hand, when the probe is completely imperfect ($\mathcal{Q}=0$)
the probe states will be indistinguishable ($|\langle
m_-|m_+\rangle|=1$) and differ at most by phase. This phase is not
contain which--way information. As a consequence the global state is
factorized. For a probe with $0<\mathcal{Q}<1$ we will have
information generation however its accessibility is only partial,
proportional to $\mathcal{Q}$. In this case the particle will
present an intermediate visibility. This means that generating the
which--way information is not enough to destroy its interference
pattern.

{\it The particle}: A general initial state of the system can be
written as $\rho_0=\{\ident+\mathbf{u}_0\cdot\mathbf{\sigma^{(1)}}\}
\otimes|m\rangle\langle m|/2 $ where,
$\mathbf{\sigma^{(1)}}=\{\sigma^{(1)}_x,\sigma^{(1)}_y,\sigma^{(1)}_z\}$
are Pauli matrices, $|\mathbf{u}_0|=1$,
$\mathbf{u}_0=\{x_0,y_0,z_0\}$ is the Bloch vector of the particle
state and $|m\rangle\langle m|$ the initial state of the probe. The
particle state immediately before the second $BS$ will be
$ \tilde{\rho}_{(1)}=\mathrm{tr}_{(2)}\{U_{F}U_{P}U_{B}\rho_{0}
U_{B}^{\dag}U_{P}^{\dag}U_{F}^{\dag}\} $
and its predictability $ \mathcal{P}_{(1)}=|x_0|$ which is identical
to the a priori predictability $\mathcal{P}_{0}$ \footnote{ The a
priori predictability is the one obtained for a interferometer
without any probe system.}. Now if the particle is detected $D_1$
and the phase difference $\Phi$ varies we will obtained an
interference fringe with visibility
$$
\mathcal{V}_{(1)}=\dfrac{I_{\textrm{max}}-I_{\textrm{min}}}{I_{\textrm{max}}+I_{\textrm{min}}}
$$
where $I_{\textrm{max}}$($I_{\textrm{min}}$) is the maximum
(minimum) fringe intensity, respectively. After secund $BS$ the
particle state will be
$ \rho_{(1)}=\mathrm{tr}_{(2)}\{U_{B}U_{F}U_{P}U_{B}\rho_{0}
U_{B}^{\dag}U_{P}^{\dag}U_{F}^{\dag}U_{B}^{\dag}\}%
$
and the intensity $I$ will be proportional to
$ I \propto \langle0|\rho_{(1)}|0\rangle=1+(z_0+iy_0)\langle
m_+|m_-\rangle e^{-i\Phi}+(z_0-iy_0)\langle m_-|m_+\rangle e^{i\Phi}
$.
This yields the visibility
$\mathcal{V}_{(1)}=\mathcal{V}_0\sqrt{1-\mathcal{D}^2}\, $, where
$\mathcal{V}_0=\sqrt{y_{0}^{2}+z_{0}^{2}\,}$ is the a priori
visibility \footnote{The a priori visibility is that obtained
without probe.}. Observe that $\,\mathcal{V}_{(1)}\,$ directly
depends on the distinguishability of the information contained in
the probe. This means that which--way information generation is a
necessary condition to destroy the interference pattern of the
particle. It is however not enough. In order to destroy interference
it is necessary {\it not only} to generate information {\it but
also} to make it accessible. The concurrence \cite{wootters2}
between the particle and the probe is given by
\begin{equation}\label{concor_1_2}
C_{(1,2)}=\mathcal{V}_0{\mathcal{D}}
\end{equation}
and it is easy to the verify that
$\mathcal{S}_{(1)}^{2}+C_{(1,2)}^{2}=1$.

{\it The probe}: From the preceding discussion its easy to see that
the complementarity relation for the probe system entangled with a
particle has the same form as that for the first system
$\mathcal{S}_{(2)}^{2}+\mathcal{C}_{(1,2)}^{2}=1$, where the duality
$\mathcal{S}_{(2)}$ must respect equation
(\ref{igualdade_dualidade}). In order to have separated expressions
to characterize the particle or wave like nature of the probe system
we will have to construct, {\it e.g.}, a quantity $\mathbb{P}$ which
reflect the particle character of the probe. Its wave like character
can then be extracted from the equality
$\mathcal{S}_{(2)}^{2}=\mathcal{S}_{(1)}^{2}$.\\
\underline{Physical conditions for constructing $\mathbb{P}$\,:}
After the probe interacts with the particle it will unitarily evolve
to $|m_{\pm}\rangle$. The operators $\Pi_\pm=|m_{\pm}\rangle\langle
m_\pm |$ are projectors and their averages
$\bigl\langle\Pi_\pm\bigr\rangle=\mathrm{tr}\bigl(\rho_{(2)}
\,\Pi_\pm\bigr)$ are the probabilities to find the probe in
$|m_{\pm}\rangle$, respectively. Similarly to the particle
predictability we would like to have $\mathbb{P}=0$ when
$\bigl\langle\Pi_\pm\bigr\rangle=1/2$ and $\mathbb{P}=1$ when
$\bigl\langle\Pi_\pm\bigr\rangle=1$. Another interesting condition
is that $\mathbb{P}$ has the interpretation of the predictability
when the probe is perfect. All these conditions will be satisfied
when we define
$\mathbb{P}=\bigl|1-2\bigl\langle\Pi_\pm\bigr\rangle\bigr|$ (and
$\pm x_0\longrightarrow |x_0|$). The probe state for the evaluation
of $\mathbb{P}$ will be
$\rho_{(2)}=\mathrm{tr}_{(1)}\{U_{B}U_{F}U_{P}U_{B}\rho_{0}
U_{B}^{\dag}U_{P}^{\dag}U_{F}^{\dag}U_{B}^{\dag}\}$. From the above
construction we get
\begin{equation}\label{pred_2}
\mathbb{P}=1-\bigl[1-\mathcal{P}_{(1)}\bigr]\mathcal{D}^2 \ \ \ \ .
\end{equation}

The factor $(1-\mathcal{P}_{(1)})$ quantifies our uncertainty as to
the path chosen by the particle. On the other hand, $\mathcal{D}^2$
measures the distinguishability of the final states of the probe.
For a value of $\mathcal{D}$ fixed, the power to predict the final
state of the probe decreases as this uncertainty grows.
The product $(1-\mathcal{P}_{(1)})\mathcal{D}^2$ quantifies the
contribution due to the wave--like character of the probe
\textit{and} the non--local correlations between the two subsystems
produced by the interaction.\\
\underline{Probe wave like character\,:} As discussed before it is
now immediate to obtain the wave--like characteristics of the probe.
One substitutes equation (\ref{pred_2}) in
(\ref{igualdade_dualidade}) and assuming that
$\mathcal{S}^{2}_{(2)}=\mathbb{P}^{2}+\mathbb{V}^{2} $\,, we get
\begin{equation}\label{visib_2}
\mathbb{V}=(1-\mathcal{P}_{(1)})\,{\mathcal{D}\,\sqrt{1-\mathcal{D}^2\,}}
\ \ \ \ .
\end{equation}
%

\begin{figure}[!htb]
\centering
  \includegraphics[scale=0.35,angle=00]{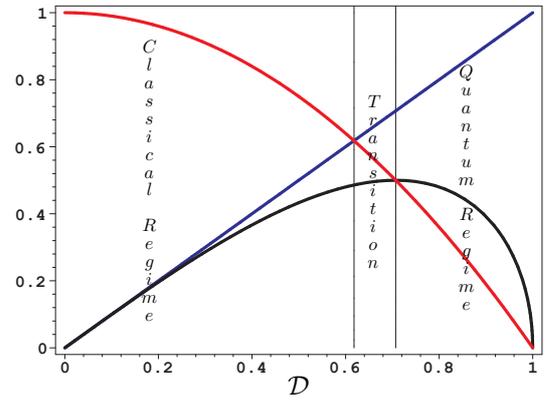}
  \caption{{\small
  {\it Probe characteristics}:
  the black (blue) curve  is $\mathbb{V}$
  ($C_{(1,2)}$), respectively, and the red curve represents
  $\mathbb{P}$ (for $\mathcal{P}_0=0$). The vertical lines locate
  the points $\mathcal{D}=1/\sqrt{2}\,$ and
  $\,{\mathcal{D}}=2/(\sqrt{5\,}+1)$. We will assume that the
  probe is within the domains of classical physics when particle
  characteristics becomes dominant
  $\mathbb{P}\geq\mathbb{V}+C_{(1,2)}$ yielding $0\leq \mathcal{D}\lesssim 0.4288$.
  It is interesting to note that in this situation the particle
  visibility will undergo a variation of $\Delta\mathcal{V}_{(1)}\approx
  0.0965$. This variation is meaningful however in experimental
  apparatuses which may reproduce these results. In realistic
  conditions the experimental error $\eta$ will exist and variations
  resulting from measurement will be accounted for by $\Delta\bar{\mathcal{V}}_{(1)}=\eta\Delta\mathcal{V}_{(1)}$.
  Depending on $\eta$ the variation $\Delta\bar{\mathcal{V}}_{(1)}$
  may be within the error bars, thus distinguishing the effects of the
  which--way information stored in the probe. This contributes
  to the idea of classical behavior see \cite{haroche} .}}\label{oprobe}
\end{figure}

Since $0\le\mathcal{D}\le 1$, the product
$\,\mathcal{D}\sqrt{1-\mathcal{D}^2}\,$ varies between 0 and $1/2$.
Two different values of $\mathcal{D}$ are associated to a particular
value of this product.
Hence,
we have $\mathbb{V}=0$ both for the perfect probe (which we
associate with the extreme quantum regime) as for imperfect one
(extreme classical regime). When the probe is perfect it entangles
with the particle and the global state of the system is an authentic
Einstein--Podolsky--Rosen pair \cite{epr}. In the opposite limit the
probe state remains unaltered and therefore no quantum features of
any sort will be present. Interesting to note that one  will always
have $\mathbb{V}\leq C_{(1,2)}$ where the equality is only reached
in trivial case $\mathbb{P}=1$.
Of course this is a particular feature of the problem in question,
since there is no local unitary operation on the probe to enhance
$\mathbb{V}$. Thus particle characteristics and its ability to
entangle will assume an important role in the complementarity
relation. Resorting to the complementarity relation for the probe,
$\mathcal{S}_{\left(2\right)}^2+ C_{\left(1,2\right)}^2=1$, and to
equation \eqref{pred_2}, it is now clear that the product
$(1-\mathcal{P}_{(1)})\mathcal{D}^2$ represents the sum of the
contributions related to the wave--like behavior of the probe,
quantified by $\mathbb{V}$, and the
non--local correlations due to the interaction, measured by $C_{\left(1,2\right)}$.\\
\underline{From the quantum to the classical limit\,:}
%
For an arbitrary particle state, the availability of the which--way
information will be minimum when $\,\mathcal{S}_{(2)}=\mathbb{P}=
1\,$, {\it i.e.}, the probe is completely imperfect and devoid of
quantum behavior. In this case, the extreme classical regime is
reached
and the probe exhibits corpuscular behavior . On the other hand, it
will be maximum when $\,\mathcal{S}_{(2)}=\mathbb{P}= |x_0|\,$ when
its behavior will be completely quantum.
Note that $\mathcal{D}=C_{(1,2)}$ when the particle is prepared in a
state which satisfies $\mathcal{P}_0=0$. In this case the probe
complementarity relation only depends on its own characteristics. We
can thus {\it define} a good probe when $\mathcal{D}>1/\sqrt{2}$,
{\it i.e.}, its imperfection will be smaller than 1/2. The opposite
limit, {\it i.e.}, $\mathcal{D}<1/ \sqrt{2}\,$ is a weak condition
since there will be a region $2/\left(1+\sqrt{5}\right) <
\mathcal{D} < 1/\sqrt{2}\,$ where $C_{(1,2)}\geq\mathbb{P}\geq
\mathbb{V}$, as shown in figure \ref{oprobe}. Thus a better
definition for a bad probe is $\mathbb{P}>C_{(1,2)} \longrightarrow
\ \mathcal{D}<2/\left(\sqrt{5}+1\right)\,$.

{\it The physics of Ramsey Zones:} To display the generality of our
findings we will discuss in their light an intriguing question.
Ramsey Zones are microwave cavities with low quality factor. It
serves the purpose of rotating effectively two level Rydberg atoms
\cite{ramsey1}. The process consists in sending atoms through the
cavity where there is a coherent field of approximately {\it one}
average photon maintained by an external source. The interesting
conceptual issue is: Why do Ramsey Zones work as if they contain
classical fields and no entanglement between atom and photon is
generated\,? A full microscopic calculation has been performed
\cite{kim}, however a convincing transparent physical picture has
not emerged. For typical values of the atomic flight velocity the
time $t_i$ of the atom field interaction is that of a few $\mu$s
\cite{haroche}. Let us consider $t_i$ sufficiently small so that one
can assume a unitary evolution during the interaction. Under this
hypothesis, the global state immediately after the interaction will
be
$$
|\phi\rangle=\dfrac{1}{\sqrt{2}}\bigl(|e\rangle|\alpha_{+}\rangle+|g\rangle|\alpha_{-}\rangle\bigr)
$$
with $|\alpha_{+}\rangle =
\sqrt{2}\biggl[\sum_{n}{C}_n\cos({\Omega}\sqrt{n+1}t_{i})|n\rangle\biggr]$
and $|\alpha_{-}\rangle =
\sqrt{2}\biggl[\sum_{n}{C}_n\sin({\Omega}\sqrt{n+1}t_{i})|n+1\rangle\biggr]$,
where ${C}_n=e^{-|\alpha|^2/2}\alpha_{}^n/\sqrt{n!}$, ${\Omega}$ is
the Rabi frequency in vacuum and $t_i$ is the necessary time for the
atom to suffer a $\pi/2$ pulse defined by
$\sum_{n}|{C}_n|^2\cos^2({\Omega}\sqrt{n+1}t_{i})=1/2$. If we have
$|\alpha|^2=1$ we will have $\,\mathbb{P}\approx0,3271\,$,
$\,\mathbb{V}\approx0,4691\,$ and $\,C_{(1,2)}\approx0,8203\,$. So
one should expect that the field in the Ramsey Zone would have
measurable quantum characteristics and behave as a good which--way
discriminator, since the information available is large
$\,\mathcal{D}\approx0,8203\,$. Why don't the expected features
appear\,?

Immediately after the atom leaves the cavity the dynamics of the
feeding source along with strong cavity dissipation act on the field
state. Since the Ramsey Zone possesses a relaxation times very
short, {\it i.e.}, smaller than $\,T_r\,=$10ns
\cite{haroche,ramsey1}, then in a time interval of the order
$\,T_r\,$ the photons inside the cavity are renewed and the state
$|\alpha\rangle$ is restored resulting $\mathcal{D}=0$ with the
information no longer available. It is important to note that this
not mean that the which--way information disappeared. It has become
stored in a system with infinite degrees of freedom which renders it
inaccessible. Therefore atom--field disentangle. In other words the
field goes almost instantaneously from a prevailing quantum regime
where it was, a priori, a good probe, to a situation where there
will be only corpuscular characteristics $\mathbb{P}=1$. This is a
classical regime as we defined. This suggests the following general
picture: systems which do not store neither make the information
about its interaction available, it can be by intrinsic properties
of the systems or by auxiliaries dynamics, may be consider classical
because the act on the quantum system without get entangled. This
might explain why quantum objects like photons cross macroscopic
apparatuses and suffer unitary evolution in the complete absence of
entanglement, {\it e.g.}: photon plus beam splitters or photons plus
quarter wave plates,  to quote some.

The present work, simple as it is, led us to believe that further
exploiting Bohr's complementarity principle may lead to very simple
and solid physical explanation for important conceptual problems
which at first sight appear to be of high mathematical intricacy.

\end{document}